\documentclass{llncs}
\pdfoutput=1

\def\papertype{techreport}

\usepackage{ifthen}
\ifthenelse{\equal{\papertype}{techreport}}
	{}
	{}
\ifthenelse{\equal{\papertype}{techreport}}
	{\newcommand{\ifthenTR}[1]{#1}}
	{\newcommand{\ifthenTR}[1]{}}
\ifthenelse{\equal{\papertype}{techreport}} 
	{}  
	{}   

\usepackage{enumitem,xcolor}

\newcounter{myprot}
\newcounter{mythm}
\newcounter{mycor}
\newcounter{myobs}
\newcounter{mydef}
\newcounter{myconj}
\usepackage[font=sf]{caption}
\usepackage{amsmath,mathtools}
\usepackage[hyphens]{url}
\usepackage{xfrac}
\usepackage{wrapfig,multirow,relsize}
\usepackage[nospace,compress]{cite}
\usepackage{microtype} 

\usepackage{xspace}

\newcommand{\para }[1]{\smallskip \noindent {\bf #1}}

\newcommand{\bc}{Bitcoin\xspace}

\usepackage{algorithmic}
\usepackage{booktabs}

\usepackage[%
rm={oldstyle=false,proportional=true},%
sf={oldstyle=false,proportional=true},%
tt={oldstyle=false,proportional=true,variable=true},%
qt=false%
]{cfr-lm}

\makeatletter
\g@addto@macro{\UrlBreaks}{\UrlOrds}
\makeatother

\makeatletter
\def\@copyrightspace{\relax}
\makeatother

\title{An Analysis of\\ Attacks on Blockchain Consensus (DRAFT)}

\author{
\hspace{-2.5ex}{George Bissias$^\dagger$, Brian  Levine$^\dagger$, A.\ Pinar Ozisik$^\dagger$, Gavin Andresen}\\
$\dagger$College of Information and Computer Sciences, Univ.\ of Massachusetts Amherst}
\institute{}
\makeatletter
\def\blfootnote{\xdef\@thefnmark{}\@footnotetext}
\makeatother

\begin{document}

\maketitle
\urlstyle{sf}

\abovedisplayskip=3pt
\belowdisplayskip=3pt
\abovedisplayshortskip=0pt
\belowdisplayshortskip=4pt

\begin{abstract}

We present and validate a novel mathematical model of the blockchain mining process and use it to conduct an economic evaluation of the double-spend attack, which is fundamental to all blockchain systems. Our analysis focuses on the {\em value} of transactions that can be secured under a conventional double-spend attack, both with and without a concurrent eclipse attack. We account for an attacker capable of increasing profits by targeting multiple merchants simultaneously. Our model quantifies the importance of several factors that determine the attack's success, including confirmation depth, attacker mining power, and a confirmation deadline set by the merchant. 
In general, the security of a transaction against a double-spend attack increases roughly
logarithmically with the depth of the block, made easier by the
increasing potential profits, but more difficult by the increasing proof of
work required.  
We find that a merchant requiring a single confirmation is protected against attackers that possess as much as 10\% of the mining power, but only provided that the total value of goods at risk for double-spend is less than 100~BTC. A merchant that requires a much longer 55 confirmations ($\approx$9 hours) will prevent an attacker from breaking even unless he possesses more than $35\%$ of the current mining power, or the value of goods at risk exceeds 1M~BTC.

\end{abstract}

\section{Introduction}\label{sec:intro}
\blfootnote{This work's copyright  is owned by the authors. A non-exclusive license to distribute the pdf has been given to arxiv.org. Last revised 2016-11-20.}
Despite the widespread adoption of blockchain-based digital currencies like Bitcoin~\cite{Nakamoto:2009}, there exists little guidance on the \emph{actual value} of goods or services that can be secured against {\em double-spend attacks} using blockchain transactions. The need for understanding the risk has always been of paramount importance to merchants; and the need is now shared by an increasing number of services that leverage blockchain transactions for settlement. For example, sidechains~\cite{Back:2014} and the Lightning Network~\cite{Poon:2015} may be deployed shortly, and the security of each depends heavily on the underlying security of the Bitcoin transactions from which they are bootstrapped. Yet, all earlier studies of the economics of double-spend attacks fall short because of the simplicity of their model and resulting inability to capture the full complexity of the problem. In the present work, we derive a novel, continuous-time model for the double-spend attack, and use it to evaluate the economics of the conventional attack; we also use it to evaluate the attack in the presence of a concurrent {\em eclipse attack}~\cite{Heilman:2015}, in which adversaries
 occlude a targeted peer's  view of the
majority's blockchain.

Double-spend attacks cannot be prevented in blockchain currencies because they are subject to the FLP impossibility result\cite{Fischer:1985}, which says, informally, that consensus cannot be reached in distributed systems that do not set a deadline for when messages (i.e., new blocks) can be received. The primary mitigating defense against
double-spends is for the merchant to wait for a transaction to receive
$z$ confirmations (i.e., $z-1$ blocks are added after the block in which
it originally appeared) before releasing goods to a customer.  Nakamoto derived the probability of success for that defense, but the result is limited because it considers
neither the cost of the attack, nor the revenue that the attacker
stands to gain.  Moreover, although it has been shown that the eclipse attack makes double-spending easier~\cite{Heilman:2015}, no prior work has yet quantified the security of a merchant's transactions in such a case where her view of the blockchain is obstructed. 

\para{Contributions.} We contribute a novel economic
evaluation of double-spend attacks in \bc that could easily be extended to similar blockchain currencies such as Litecoin\cite{litecoin} or Zerocash~\cite{sasson:2014}.  We derive and validate equations for the value of transactions that can be secured against a double-spending adversary, who controls any portion of the mining power less than a majority. We also evaluate the double-spend attack
when conducted contemporaneously with the eclipse attack.

Our results quantify Bitcoin's security as a currency. We show that the correct
attacker model considers not just the attacker's mining power and transaction confirmation depth $z$, but also the attacker's potential reward, or \emph{goods at risk}, which can be conservatively estimated by a merchant as the summed value of coin in the $z$ confirmation blocks that is exchanged between individuals (turnover)\cite{meiklejohn:2013,ron:2013}.
We find that blockchain security against double-spend attacks increases roughly
logarithmically with block depth, made easier by increasing goods at risk, but more difficult by the increasing proof of work required. Our model also quantifies the synergistic value of a concurrent eclipse attack.

For example, when the summed turnover for a transaction's first confirmation block is as high as 100~BTC, we determine that a single confirmation is protected against attackers that can increase the current mining power by no more than $10\%$. Waiting for significantly more confirmations increases a transaction's security considerably. With 55 confirmations and aggregate turnover of up to 1M~BTC, a transaction can be protected from a double-spend attack as long as the attacker possesses less than $35\%$ of the current mining power. We also demonstrate quantifiably that if merchants impose a conservative confirmation deadline, then a concurrent eclipse attack can only make double-spends more profitable if the attacker possesses less than 35\% of the mining power or if the merchant requires fewer than 10 confirmations.

\section{Background and Related Work}\label{sec:background}

Using a blockchain as a method for distributed consensus was first proposed by Nakamoto as part of his or her development of the Bitcoin digital currency~\cite{Nakamoto:2009}. Blockchains allow for an open group of peers to reach consensus, while mitigating Sybil attacks\cite{Douceur:2002} and the limitations imposed by the FLP impossibility result~\cite{Fischer:1985} through a {\em mining} process. We refer the uninitiated reader to Appendix~\ref{sec:backappendix} for a detailed overview of blockchain consensus and Bitcoin. Additionally, several articles are available that offer summaries of broader \bc research issues\cite{Tschorsch:2016,bonneau:2015,Croman:2016}. 
Below, we summarize two particularly relevant attacks, and then summarize why our contributions are distinct from related work. 

\subsection{Relevant Attacks} 
\para{Double spending.} 
A fundamental attack against \bc is the {\em double-spend} attack~\cite{Nakamoto:2009},
which works as follows. An attacker creates a transaction that moves
funds to a merchant's address. After the transaction appears in the
newest block on the main branch, the attacker takes possession of the purchased
goods. Using his mining power, the attacker then immediately releases
two blocks, with a transaction in the first that moves the funds to a
second attacker-owned address. Now the attacker has the goods and his
coin back. To defend against the attack, a merchant can refuse to
release goods to a Bitcoin-paying customer until $z$ blocks have been
added to the blockchain including the first block containing a
transaction moving coin to the merchant's address.  Nakamoto
calculated the probability of the attack succeeding assuming that the
miner controlled a given fraction of the mining power~\cite{Nakamoto:2009};
for a given fraction, the probability of success decreases exponentially as $z$
increases. 

In general, a merchant may wait $z$ blocks before releasing goods,
which can thwart an attacker.
But choosing the minimum value of $z$ that secures a transaction is an
unresolved issue. The core \bc client shows that a transaction is
unconfirmed until it is 6 blocks deep in the
blockchain\cite{bitcoin:confirmation}, and the advice from researchers to
policymakers can be vague; e.g., ``for very large transactions, coin
owners might want to wait for a larger number of block
confirmations''~\cite{Bonneau:2015a}.   

\para{Eclipse attacks.}
Heilman et al.\ showed that Bitcoin's p2p network peer discovery
mechanism is vulnerable to eclipse attacks~\cite{Heilman:2015}, which
occlude a victim peer's view of the blockchain.  For example, if an
adversary controls a botnet, he can fill a peer's table of possible
neighbors, resulting in a very high chance the victim will connect
only to the attacker.  Alternatively, the eclipse can involve
controlling a victim's local connection to the Internet.

Heilman et al.\ also showed that eclipse attacks can be used as a tool
to increase the effectiveness of the double-spend attack on a
merchant. First, the attacker eclipses the merchant's view of the
blockchain. Then, he sends the merchant a seemingly honest transaction
$\cal{H}$, which contains the payment for a good. Third, the attacker
sends to the miners a faulty transaction $\cal{F}$ that moves the
funds elsewhere.  Next, he creates and sends a series of $z$ blocks to
the victim merchant such that $\cal{H}$ is part of the first block.
Finally, he continues the eclipse until the real blockchain has
progressed by at least $z+1$ blocks. At that point, he has both the
goods and the Bitcoin that the merchant intended to keep.

\subsection{Related Work}\label{sec:related}
Nakamoto derived the double-spend attack's success probability in the original Bitcoin paper~\cite{Nakamoto:2009}. The main limitation of this approach is that it fails to include attack cost, which would place the severity of an attack in a real world context. Moreover, the model itself is also overly simplified in that it models the creation of an entire sequence of blocks as a single Poisson process. Accordingly, we cannot rely on the accuracy of the model for large numbers of consecutive blocks, which is necessary for determining the security of high-value transactions.

In their paper introducing the GHOST protocol~\cite{sompolinsky:2015}, Sompolinsky et al. extended Nakamoto's model by incorporating network delays and by allowing the expected block creation time to deviate from the 10-minute average used today. With this model, they showed that double-spend attacks become more effective as either the block size or block creation rate increase (when GHOST is not used). With a trivial change, our work could also vary the expected block creation time. A less trivial (but quite interesting) change would allow us to also model network delays.

Sapirshtein et al.~\cite{Sapirshtein:2015} first observed that some double-spend attacks can be carried out essentially cost-free in the presence of a concurrent selfish mining~\cite{eyal:2014} attack. More recent work extends the scope of double-spends that can benefit from selfish mining to cases where the attacker is capable of \emph{pre-mining} blocks on a secret branch at little or no opportunity cost~\cite{Sompolinsky:2016} and possibly also under a concurrent eclipse attack~\cite{Gervais:2016}. 
The papers identify the optimal mining strategy for an attacker and quantify the advantage that he can expect to have over the merchant in terms of pre-mined blocks.
This analysis is complementary to ours; it is possible to relatively easily incorporate the pre-mining advantage into our model by simply changing the attacker's block target from $z$ to $z-c$. We note that pre-mining in the context of the eclipse attack may not be feasible since an eclipse cannot generally be carried out for an indefinite period of time. Nevertheless, we intend to update both of our double-spend analyses to account for cost-free pre-mining in future work.

The objective of Rosenfeld~\cite{Rosenfeld:2012} is most similar to ours and his analysis is a great improvement over the 6-block rule. However, his model cannot be applied to the concurrent eclipse attack scenario and he makes several simplifying assumptions that render the results for the conventional double-spend attack less  accurate than ours, particularly as the number of required confirmations, $z$, grows (we compare quantitatively in Fig.~\ref{fig:min-q-static-v}). Additionally, his approach --- as well as all of those cited above --- models only the \emph{order} of block creation and not block mining time explicitly. As a result, it is difficult to extend his results to model cost in circumstances where the attacker is given a specific deadline (as we have done in our eclipse attack analysis) or where an attacker drops out because the honest miners have already won. We develop a richer, continuous-time model that explicitly accounts
for attacker cost as a function of mining duration.

Heilman et al.\ offered a detailed analysis of the mechanics of an
eclipse attack, as well as several protocol-level defenses to the
attack.  But they attempted no analysis of an attacker's economic incentives.  As a result, it remains unclear what minimum number of
confirmations, $z$, are sufficient to secure a given value of
purchased goods. In this paper, we derive a model for the profit
received by an adversary who launches an eclipse attack, and use it to
determine the attacker's break-even point for various values of $z$.

\newcommand{\zpo}{z\!\!+\!\!1}

\section{Analysis of Double-spend and Eclipse Attacks}\label{sec:eclipse}

In this section, we compute the security of a transaction against a double-spend attack, both with and without a concurrent eclipse attack, in cases where the transacted coin is used as payment for goods or services, which we call \emph{goods at risk}. Specifically, we are able to determine the fraction of mining power $q$ required by an attacker to profitably double-spend a transaction given the value of goods at risk $v$ and transaction confirmation blocks $z$ required by the target (merchant, service, etc.).  As we explain below, $v$ is conservatively estimated as not just the Bitcoin value of goods sold by a single merchant, but as the sum amount of coins transferred between entities (turnover) by all transactions in the $z$ confirmation blocks.  Our guiding principle is that \emph{a resource is secure from an
  attack only if it is worth less than the attack's cost}.  We find that because of the
well-behaved statistical properties of mining times and transparency
of transaction values, Bitcoin is particularly amenable to such an
analysis.

\para{Attack cases.}
In this section, we analyze two double-spend attack strategies. \emph{Case 1} assumes that the attacker is capable of eclipsing the merchant while conducting a double-spend attack. \emph{Case 2}  assumes that the attacker elects not to employ an eclipse attack (or equivalently, fails in an attempt to do so). Assuming that the merchant follows our mitigation guidelines, we provide bounds on the expected break-even point for the attacker in both cases. 
We find that there are two distinct attack regimes where one case dominates the other based on the attacker's share of the mining power $q$. When $q \leq 0.35$ and $z$ remains relatively low, Case 1 affords the attacker a lower break-even point. The opposite is true for $q \geq 0.35$ and large $z$, where Case 2 dominates.

\para{Attack target.} Throughout most of this section we proceed under the assumption that the attacker targets a single merchant, which is not strictly true. First, the target need not be an individual at all; the attacker could instead target blockchain-based services. Second, he could exploit multiple targets simultaneously. We discus the ramifications of the former in Section~\ref{sec:discussion} and the latter presently. Simultaneous attacks increase the attacker's potential profit by allowing for multiple double-spends to be placed in a single block. Moreover, even if a merchant requires $z$ confirmations, the attacker might be lucky to find other merchants requiring only one, which means that the attacker could potentially profit from all $z$ blocks. The potential profit in a block can be bounded by the sum of all Bitcoin transferred, or \emph{turned over}, between distinct entities. We call the aggregate turnover value for $z$ blocks the \emph{$z$-maximal goods} because it represents the maximum Bitcoin value of goods that could be exploited by the attacker in the $z$ confirmation blocks. Several past works have evaluated metrics for estimating that value\cite{ron:2013,meiklejohn:2013}. Thus, we can model the simultaneous attack scenario by letting the goods at risk, $v$, be equal to the $z$-maximal goods.

\para{Mitigation measures.} We assume that the merchant will take certain precautions. First, because the attacker is capable of profiting from the $z$-maximal goods, the merchant will set $z$ adaptively based on the turnover of each block as it is added to the blockchain. Specifically, she will only release goods after $z$ confirmations if the expected cost to an attacker exceeds the $z$-maximal goods. Second, when the merchant is concerned about a simultaneous eclipse attack, she will impose a deadline $d$ for the receipt of all $z$ confirmations. It is always possible to double-spend against an eclipsed merchant --- only the attack duration varies based on the attacker's fraction of mining power $q$ --- thus $d$ serves to increase the attacker's cost by increasing his loss rate. Because $z$ is set dynamically, $d$ would most naturally be defined as a linear function of $z$.

\subsection{Attacker model}

We make the following simplifying assumptions about the attack environment as well as the attacker's capabilities and behaviors.

\begin{itemize}[itemsep=3pt,topsep=1pt]
\item The attacker's mining power constitutes a fraction $0<q<0.5$ of the total mining power. When $q \geq 0.5$, the attacker holds a majority of the mining power, in which case Bitcoin cannot secure any transaction. 

\item The network is correctly calibrated so that a block is
produced roughly once every 10 minutes given the current mining power, which
is generally true in the real system. Bitcoin adjusts its difficulty
once every 2,016 blocks (about every two weeks), and we assume these
attacks have no affect on the difficulty while they are run. 

\item The eclipse attack succeeds without fail.  An eclipse
attack is likely to incur some cost, but we do not include it because
that cost is hard to estimate. For example, in the most general
scenario, a botnet might be required~\cite{Heilman:2015}. On the other
hand, if a merchant is physically accessible and has only a single,
unsecured wireless link to the Internet, eclipse attacks are much
simpler and much less costly.  We also assume the attacker does not
launch a denial-of-service attack on honest miners. 

\end{itemize}

\para{Attacker strategy Case 1:} The attacker launches an eclipse attack against one or more merchants. He diverts his mining power, $q$, from the main branch to mining an alternate \emph{fraudulent branch} that contains in the earliest block  a transaction that moves coins to the merchant; the blocks are sent to the merchant as they are mined. Once the attacker reaches $z$ blocks on the fraudulent branch, the merchant releases the goods to the attacker. The attacker ceases to mine on the fraudulent branch, and waits until the main branch grows longer than the fraudulent branch. He then ceases to eclipse the merchant, allowing her to realize that the actual longest branch does not contain the transaction that transferred coin to her.

In principle, the attacker has an unlimited amount of time to produce the $z$ blocks he uses to unlock the goods from the merchant. He is limited only by the amount of time he is able to eclipse the merchant and the time his is willing to divert his mining power away from the main branch. However, the merchant will likely suspect that she is being eclipsed if the actual time it takes for her to receive $z$ blocks is drastically different than the expected time (roughly 10 minutes per block). Therefore, we propose that the merchant refuse to hand over the goods if the $z$ blocks are not received by her \emph{deadline} $d$, a parameter that we described in our discussion of mitigation measures. Given this change in policy, the attacker will naturally adjust to cease mining on the fraudulent branch once he either mines $z$ blocks or the deadline has passed. For this case, we do not consider the possibility that the attacker attempts to replace the main branch with his false branch, thus we assume that the coinbase rewards earned on the fraudulent branch are useless.

\para{Attacker strategy Case 2:} No eclipse attack is leveraged, but the attacker again possesses fraction $q$ of the total mining power.  This time, after releasing the transaction that pays coin to the merchant, he races to mine $z+1$ blocks on a \emph{secret} fraudulent branch that does not contain the transaction in any of its blocks. Meanwhile, the rest of the miners have picked up the payment transaction and added it to the main branch on which they mine. The merchant will release the goods once the main branch reaches length $z$. The attacker does not release any of his blocks until the fraudulent branch has reached length $z+1$, at which point he releases all block simultaneously. If the fraudulent branch is longer than the main branch, then the attacker will have successfully double-spent the coins that he originally transferred to the merchant. 

A persistent attacker could conceivably continue to mine indefinitely, even after the main branch far surpasses the fraudulent branch. There will remain, in any case, a non-zero probability of success. However, one of Nakamoto's fundamental results is that there will be an exponentially diminishing probability of success with every block that the attacker falls behind~\cite{Nakamoto:2009}. Moreover, the cumulative cost will eventually become prohibitive to any rational attacker. Therefore, we assume the attacker will eventually quit if he remains behind for a sufficiently long period of time. Determining the optimal drop out point is left for future work; here we arbitrarily assume that the attacker will drop out if the main branch mines $z+1$ blocks before he can. This choice has the desirable property (for the attacker) that the expected cost for any outcome will not exceed the cost he was willing to incur for a successful attempt. 

\subsection{Case 1 Analysis: Eclipse-Based Double-Spend}
\label{sec:case1}
We determine \bc's security with respect to  an eclipse-based
double-spend attack by quantifying a potential attacker's economic
break-even point. Break-even occurs when  revenue
$R$ less  cost $C$ is zero:
\begin{align}
R-C=0.
\end{align}
We assume an attacker has fraction $q$ of  total mining power, and the merchant will not release goods until a paying transaction is $z$-blocks deep in the main chain.  If the $z$th  block has not been announced by $d$ minutes, the sale is nullified.

Let $X^q_i$ be a random variable representing the time it takes for the
attacker to mine the $i$th block using mining power $q$, and let
	\begin{align}
	X = \sum_{i=1}^z X_i^q.
	\end{align}
Here, $X$ represents the time it takes the attacker to reach $z$ blocks using mining power $q$.
Define $C(x; d, q)$ to be the cost of an attack with duration $x$ minutes
and deadline $d$ that uses mining power $q$.  To calculate cost, we
assume that the miner will stop mining once he mines the $z$ blocks,
but will continue to mine until the deadline if he is unsuccessful.
Cost can be measured in terms of the
opportunity cost for diverting the mining power from performing honest
mining, that is, the attacker could have earned the
block reward of $B$ from the main branch.  Blocks are
mined, we expect, every 10 minutes. Therefore
\begin{eqnarray}
  C(x; d, q) = \left \{ \begin{array}{rl}
                          \frac{qxB}{10}, & x \leq d \\
                          \frac{qdB}{10}, & x > d
                        \end{array}
                                           \right. .
\label{eqn:case1_cost}
\end{eqnarray}
Mining is an example of a Poisson process because, under constant
mining power, blocks are mined continuously and independently at a
constant average rate.  Therefore
$X_i^q \sim \texttt{exponential}(\beta)$ with $\beta = 10 / q$.  It is
well known that the exponential distribution is a special case of the
gamma distribution with shape parameter $\alpha=1$.  Furthermore, the
sum of $z$ gamma distributions with shape $\alpha=1$ and the same rate
$\beta$ is again gamma with rate $\beta$ and shape $z$.  Thus
$X \sim \texttt{gamma}(z, 10/q)$.  Let $g(x; \alpha, \beta)$ be the
density function for the distribution $\texttt{gamma}(\alpha, \beta)$,
and let $G(x; \alpha, \beta)$ be the CDF.  It follows that
\begin{eqnarray}
  E[C(X; d, q)] &=&  \int_0^{\infty} C(x;d, q) g(x, z, \beta) dx  \notag\\
	& = & \int_0^{\infty} C(x; d, q) \frac{1}{\beta^z (z-1)!} x^{z-1} e^{-x / \beta} dx \notag\\
	& = & \frac{qB}{10}  \left [ \int_0^d \frac{1}{\beta^z (z-1)!} x^z e^{-x / \beta} dx  + d (1 - G(d, z, \beta)) \right ]\notag \\
	& = &  \frac{qB}{10}  \left [ \beta z \int_0^d \frac{1}{\beta^{(z+1)} z!} x^z e^{-x / \beta} dx  + d (1 - G(d, z, \beta)) \right ]\notag \\
	& = &   \frac{qB}{10}  \left [ \frac{10z}{q} G\left(d; z+1, \frac{10}{q}\right)  + d \left(1 - G\left(d; z, \frac{10}{q}\right) \right) \right ] \notag\\
    & = & \frac{qdB}{10}  + zB G(d; \zpo, \sfrac{10}{q}) - \frac{qdB}{10} G(d; z, \sfrac{10}{q}).
          \label{eq:ec-min-cost}
\end{eqnarray}
Consider now the attacker's revenue $R(x;d)$.  If he succeeds in the
attack, then he will earn revenue $v$ and will earn nothing otherwise.
Formally,
\begin{equation}
R(x; d) = \left \{
\begin{array}{rl}
v, & x < d \\
0, & x \geq d
\end{array}
\right . .
\end{equation}
The probability of his success is
\begin{eqnarray}
P(X \leq d) &=& G(d; z, 10/q).\label{eq:c1-pxd}
\end{eqnarray}
Hence the expected revenue is given by
\begin{eqnarray}
E[R(X;d)] &=& v  G(d; z, 10/q).
\end{eqnarray}
Using the fact that expected break-even occurs when
$E[R(X; d)] - E[C(X; d, q)] = 0$ we have our {\bf result for Case 1}:
\begin{eqnarray}
 v &=&  \frac{E[R(X;d)]}{G(d; z, 10/q)}\notag\\
 &=&  \frac{E[C(X;d, q)]}{G(d; z, 10/q)}\notag\\
&=& \frac{\frac{qdB}{10}  + zB G\left(d; \zpo, \frac{10}{q}\right) - \frac{qdB}{10} G\left(d; z, \frac{10}{q}\right)}{G(d; z, 10/q)} \notag\\
&=& \frac{\frac{qdB}{10}  + zB G\left(d; \zpo, \sfrac{10}{q}\right)}{G(d; z, \sfrac{10}{q})}- \frac{qdB}{10}.\label{eq:c1-v}
\end{eqnarray}

\subsection{Case 2 Analysis: Double-Spend Without The Eclipse Attack}

In this case, we assume that an eclipse attack is not employed by the attacker. 
By comparing results to Case 1, we can determine which is the more effective attacker strategy. 
Recall that the attacker builds a \emph{fraudulent branch} holding the double-spend transaction, and  honest miners build the \emph{main branch} holding the payment transaction. The attack succeeds if the
fraudulent branch becomes the
main branch. No deadline is enforced by the merchant.
She does, however, enforce an embargo on goods until the payment transaction
is $z$-blocks deep.

Let $Y_i^q$ be a random variable denoting the time it takes the attacker to mine the $i$th block if he controls fraction $q$ of the total mining power.  Similarly, define $M_i^q$ to be the time
it takes the honest miners to mine block $i$ given that they control fraction
$1-q$ of the mining power.  Finally, define
	\begin{align}
	Y = \sum_{i=1}^{z+1} Y_i^q
	\end{align}
	and 
	\begin{align}
	M = \sum_{i=1}^{z+1} M_i^q
\end{align}
 to be
the time it takes for the attacker and other miners, respectively, to
mine $z+1$ blocks.  For the attack to be a success, it must
be the case that $Y < M$.  We assume that the attacker will stop
mining when he reaches $z+1$ blocks on the fraudulent branch or when
the honest miners reach $z+1$ blocks on the main branch, whichever
happens first.  

Analogously to Section~\ref{sec:case1}, we define
$C(y,m; q)$ as the cost to the attacker when he possesses mining power $q$ and the attacker and honest miners each mine for $y$ and $m$ minutes respectively:
\begin{eqnarray}
C(y, m; q) = \left \{ \begin{array}{rl}
\frac{qyB}{10}, & y \leq m \\
\frac{qmB}{10}, & y > m
\end{array}
\right. .
\end{eqnarray}
Just like $X$ in Section~\ref{sec:case1}, both $Y$ and $M$ have gamma
distributions, this time with rate parameters $\beta_Y = 10/q$ and
$\beta_M = 10/(1-q)$, respectively.  Specifically,
$Y \sim \texttt{gamma}(z+1, \beta_Y)$ and
$M \sim \texttt{gamma}(z+1, \beta_M)$.  Define $g(x;\alpha, \beta)$
and $G(x; \alpha, \beta)$ as in Section~\ref{sec:case1}.  It follows
that
\begin{eqnarray}
E[C(Y, M; q)]
&=&  \int_0^{\infty}\!\!\!\int_0^{\infty} \!\!\!C(y, m; q) ~ g(m; \zpo, \beta_M) g(y; \zpo, \beta_Y) ~ dy ~ dm \notag 
\\	&=&  \int_0^{\infty} g(m; \zpo, \beta_M)  \biggl( \frac{qB}{10} \int_0^m y g(y; \zpo, \beta_Y) dy~+\notag\\
 		&&\hspace{1in}\frac{qmB}{10} \int_m^{\infty} g(y; \zpo, \beta_Y) dy \biggr) dm \notag\\
	&=&   \frac{qB}{10} \biggl( \int_0^{\infty} g(m; z+1, \beta_M) \int_0^m y g(y; z+1, \beta_Y) ~ dy ~ dm +\notag\\
		&&\hspace{5ex} \int_0^{\infty} m g(m; z+1, \beta_M) \int_m^{\infty} g(y; z+1, \beta_Y) ~ dy ~ dm \biggr)\notag
\end{eqnarray}
\begin{eqnarray}
&=&  \frac{qB}{10} \biggl( \int_0^{\infty} g(m; z+1, \beta_M) \int_0^m \beta_Y (z+1) g(y; z+2, \beta_Y) ~ dy ~ dm + \notag\\
	    && \int_0^{\infty} \beta_M (z+1) g(m; z+2, \beta_M) \int_m^{\infty} g(y; z+1, \beta_Y) ~ dy ~ dm \biggr)\notag \\
	&=&  \frac{q B(z+1)}{10} \biggl(\beta_Y \int_0^{\infty} g(m; z, \beta_M) G(m; z+2, \beta_Y) ~ dm +\notag	\\
	     && \beta_M  \int_0^{\infty} g(m; z+1, \beta_M) \left( 1 - G(m; z+1, \beta_Y) \right) ~ dm \biggr)\notag \\
&=& B(z+1)\biggl( \int_0^{\infty}\!\!\!g(m; z, 10/(1-q)) G(m; z+2, 10/q) ~ dm +\notag\\
&&\frac{q}{1-q}  \int_0^{\infty}\!\!\!g(m; z+1, 10/(1-q))\!\left( 1 - G(m; z+1, 10/q) \biggr)  dm \right).
\label{eqn:case2_cost}
\end{eqnarray}
\begin{figure*}[t]
\centerline{\includegraphics[width=\columnwidth]{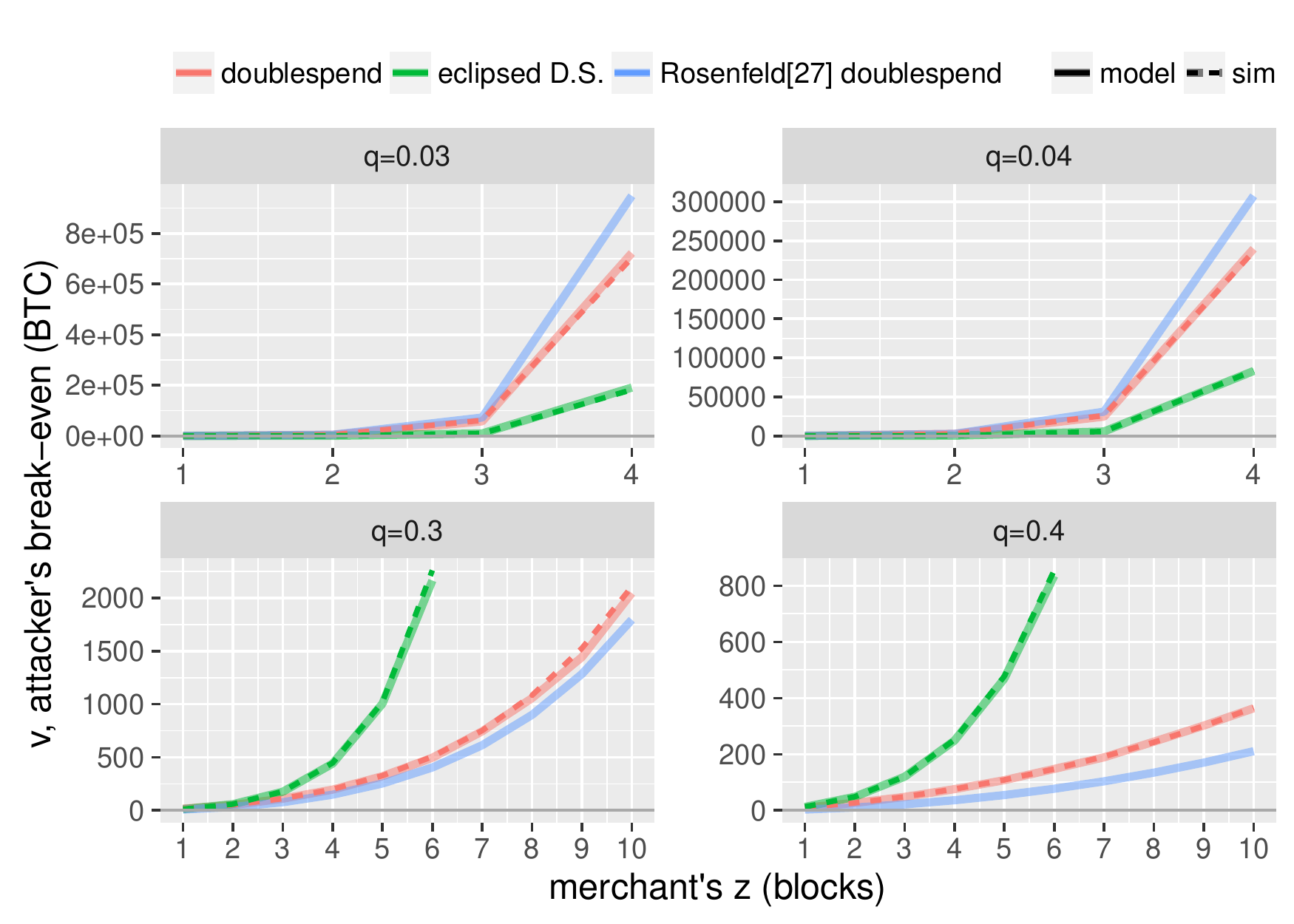}} 
 \caption{For an attacker with mining power $q$, these plots show 
the value of goods at risk $v$ required to expect to break even given the merchant's confirmation requirement $z$. An eclipse-based double-spend attack 
(Case 1, Eq.~\ref{eq:c1-v}) appears in green; a standard double-spend attack (Case 2, Eq.~\ref{eq:c2-v}) appears in red. For the eclipse attack we 
set the deadline to $d=10z$ minutes. An alternative visualization of the equations appears in the appendix, Figure~\ref{fig:log}. 
 Results from our model are 
shown as solid lines; results from an independent Monte Carlo 
simulation  are shown as a dashed line.  Rosenfeld's model~\cite{Rosenfeld:2012} of Case 2 is shown in blue.}
\label{fig:econ-z-wait-limited}
\hspace{-1cm}
\end{figure*}
To determine the attacker's expected break-even point, we must
also calculate his expected revenue.  For any given $z$, the
attacker's revenue when it took him $y$ minutes to mine $z+1$ blocks
on the fraudulent branch while it took the other miners $m$ minutes to
mine $z+1$ blocks on the main branch is given by
\begin{equation}
R(y, m; z) = \left \{
\begin{array}{rl}
v + (z+1)B, & y < m \\
0, & y \geq m
\end{array}
\right . .
\end{equation}
Revenue differs from that collected in Case 1 because the successful attacker will earn the coinbase reward for each block he mines.  The probability of
attack success is equal to
\begin{eqnarray}
P[Y < M]&=&\int_0^{\infty}\!\!\!\!\int_0^m\!\!\!g(m; \zpo, 10/(1-q))g(y; \zpo, 10/q) ~ dy ~ dm \notag\\
&=& \int_0^{\infty}\!\!\! g(m; \zpo, 10/(1-q)) G(m; \zpo, 10/q) ~ dm .\label{eq:P}
\end{eqnarray}
Therefore, the expected revenue can be calculated as
\begin{eqnarray}
E[R(Y, M; z)] &=&
\int_0^{\infty}\!\!\!\int_0^m\!\!(v + (\zpo)B)  g(m; \zpo, 10/(1-q)) g(y; \zpo, 10/q) dy~dm \notag \\
\ifthenelse{\equal{\papertype}{techreport}}{
&=&(v + (\zpo)B) \int_0^{\infty}\!\!\! g(m; \zpo, 10/(1-q)) G(m; \zpo, 10/q) dm \notag \\
}{}
&=&(v + (\zpo)B) ~ P[Y < M].
\end{eqnarray}
The expected break-even point is
value of $v$ for which revenue minus cost is zero.
\begin{align}
 &E[R(Y, M; z)] - E[C(Y, M; q)] = 0.   \\ 
\intertext{Substituting for revenue,}
 & (v + (\zpo)B) P[Y < M] =  E[C(Y, M; q)],&  \\ 
 \intertext{and rearranging, we have our {\bf  result for Case 2} as follows:}
\ifthenelse{\equal{\papertype}{techreport}}{
	v =&  \dfrac{E[C(Y, M; q)] - (z+1)B P[Y < M]}{P[Y < M]} \notag\\
	}{}
v =&  \dfrac{E[C(Y, M; q)] }{P[Y < M]}- (z+1)B. \label{eq:c2-v}
\end{align}
Note that Eq.~\ref{eq:c2-v} is fully expressed by substituting  Eqs.~\ref{eqn:case2_cost} and~\ref{eq:P}.

\begin{figure}[t]  
\begin{minipage}{.6\columnwidth}
\centerline{\includegraphics[width=.9\columnwidth]{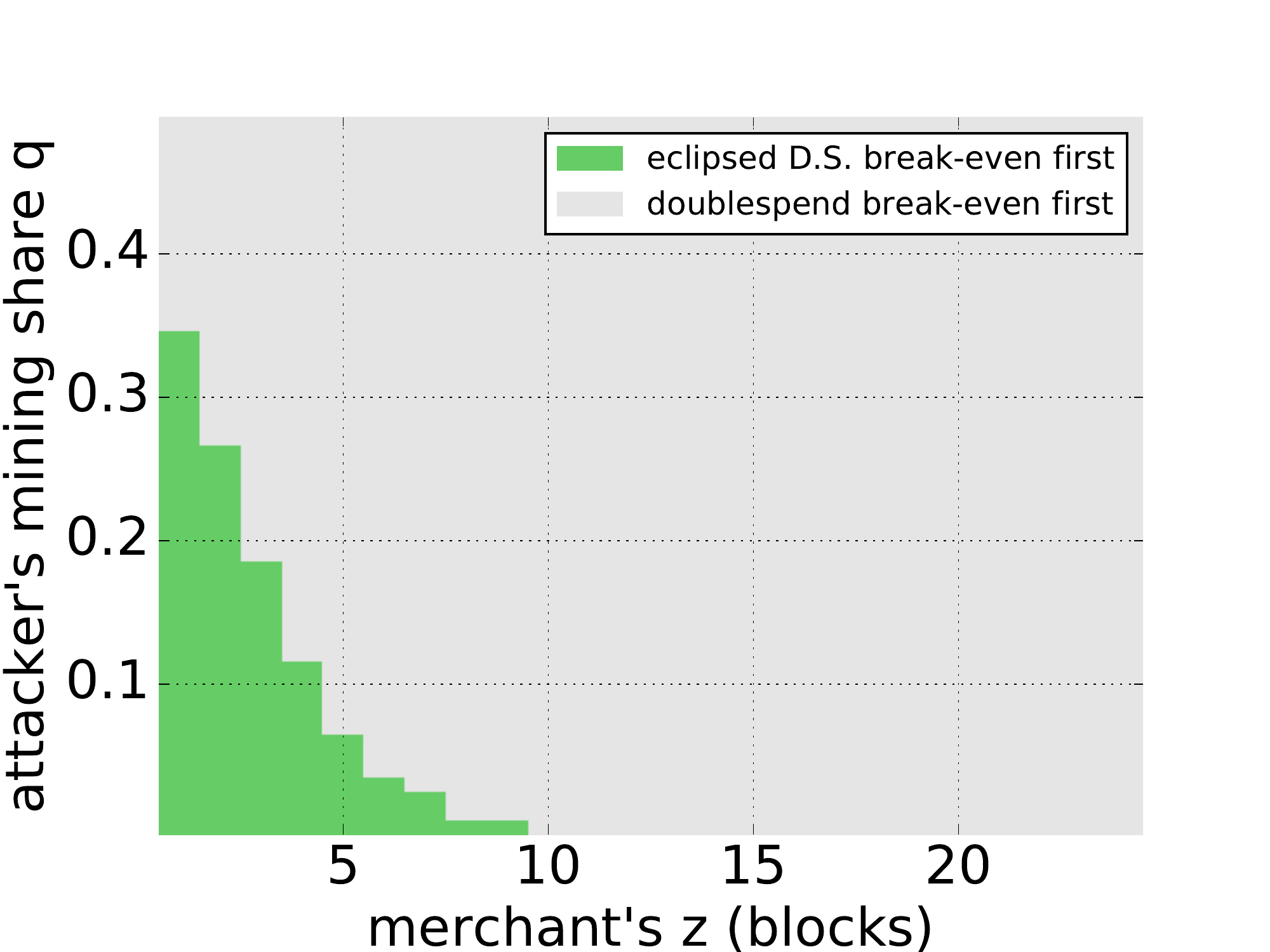}} 
\end{minipage}\hfill
\begin{minipage}[c]{.4\columnwidth}
\caption{Regimes where one attack case breaks even before the other. The green region denotes combinations of $z$ and $q$ where Case~1, the concurrent eclipse and double-spend attack, will realize a profit for lower values $v$ than in Case~2. The grey region indicates the opposite: Case~1 profits with lower $v$ than Case~2.}  \label{fig:heatmap}
\end{minipage}
 \end{figure}

\section{Discussion}
\label{sec:discussion}

In this section, we use the derivations from the previous section to quantitatively compare attacker strategies and discuss implications for blockchain systems and their users. In addition to the assumptions outlined in Section~\ref{sec:eclipse}, we fix the deadline at $d=10z$ for Case 2, which implies that the merchant will release the goods only if the payment transaction receives $z$ confirmations within $10z$ minutes. We discuss the practical ramifications of this choice later in this section.

\para{Accuracy of model.} Fig.~\ref{fig:econ-z-wait-limited}
compares Eqs.~\ref{eq:c1-v} and~\ref{eq:c2-v}, the break-even value for
a rational attacker performing a double-spend attack with and without
eclipsing the merchant, respectively. 
In general, the break-even point increases with $z$, and for lower values of $q$ the break-even point grows particularly rapidly. Thus we limit $z\leq 4$  for small $q$ in order to more easily discern the differences between curves on the same axes.  An alternative visualization of the equations appears in the appendix, Fig.~\ref{fig:log}.
Fig.~\ref{fig:econ-z-wait-limited} also includes the results of an independent Monte Carlo (MC) simulation of both attacks, executed thousands of times for each point and is in very close agreement with our model.

\para{Better to eclipse or not?} Whether using an eclipse attack is an advantage to the attacker depends on both $q$ and $z$. Fig.~\ref{fig:heatmap} shows that the concurrent eclipse attack of Case 1 affords the attacker with a lower break-even point (which is advantageous) than Case 2 for relatively low values of $z$; the advantage holds for fewer values of $z$ as $q$ increases. Indeed, once $q>0.35$ or $z \geq 10$, Case 2 always offers a lower break-even point than Case 1. The precise regime where one case dominates the other depends on the choice of deadline $d$ in Case 1. A longer deadline will decrease the cost for the attacker, which will lower his break-even point, and tend to enlarge the green region in Fig.~\ref{fig:heatmap}. Nevertheless, we expect that Case 2 will always dominate Case 1 for large $z$ because the fraudulent chain replaces the main chain in the latter case, earning the attacker the cumulative block reward.
 
 \begin{figure}[t]  
\begin{minipage}{.73\columnwidth}
\centerline{\includegraphics[width=\columnwidth]{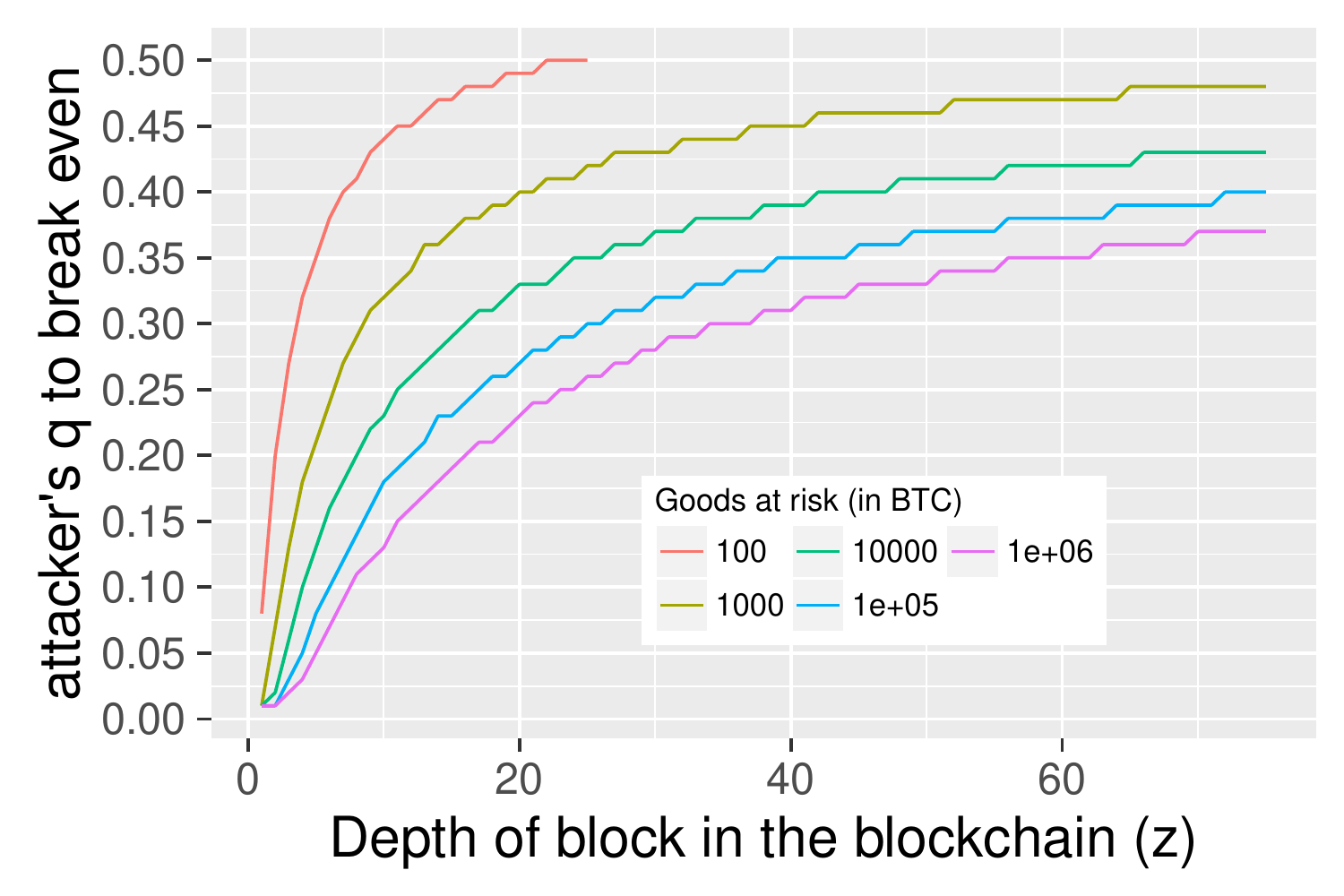}} 
\end{minipage}\hfill
\begin{minipage}[c]{.26\columnwidth}
\caption{Minimum mining power  $q$ required of attacker  to break-even for given merchant confirmation requirement $z$ and
  goods at risk $v$. For each value of $z$ and $v$,
  the most effective double-spend attack is used (eclipse or
  not-eclipse based) to find the lowest value $q$ that allows the
  attacker to break even.  }  \label{fig:min-q-static-v}
\end{minipage}
 \end{figure}

\para{Comparison to Rosenfeld\cite{Rosenfeld:2012}}. Rosenfeld offers a model for $v$ that is directly comparable to what we call Case 2 (double-spend without the eclipse attack), but his model is not capable of addressing Case 1. In terms of our notation, his model yields the bound $v>(1-r)zB/r$, where $r$ is a discrete model\footnote{Specifically, Eq.~1 in \cite{Rosenfeld:2012}.} of the attacker's probability of success given $z$ and $q$. Fig.~\ref{fig:econ-z-wait-limited} plots his model next to ours as well as our independent Monte Carlo simulation results. His model provides a reasonable fit for $q=0.3$, but is not accurate otherwise; the error in Rosenfeld's model tends to increase with $z$. For example, when $q=0.04$ and $z=4$, his model over-shoots the MC results by almost 35\%,
 and when $q=0.4$ and $z=10$ his model underestimates the true break-even point by over 40\%.

\para{Revenue required to break even.} In general, double-spending
attacks are more efficient with higher mining power $q$ and goods at risk $v$. Higher $q$ means less risk of losing, and higher $v$ means greater potential profit.  
Each curve in Fig.~\ref{fig:min-q-static-v} represents a single value for goods at risk and shows how the mining power necessary for the attacker to break even varies with $z$\footnote{In Fig.~\ref{fig:min-q-static-v}, Case~1 values are easily generated directly from Eq.~\ref{eq:c1-v}; the equation contains no integrals and most scientific software packages can deliver values from the \texttt{gamma} distribution. In contrast, Case~2 involves a somewhat difficult integration (with no analytical solution), and numerical integration packages fail for portions of our parameter space. We instead used {\em Monte Carlo integration} --- a distinct technique from the Monte Carlo simulation discussed above ---  to generate points from Eq.~\ref{eq:c2-v}}. 
For each pair of values $v$ and $z$, the minimum $q$ value from the most
effective double-spend attack strategy is reported in the plot (the case that
breaks even for lowest value $q$).

From the plot we can see that, for
low values of $z$, even attackers with limited mining power can break even for low values of goods at risk. On the other hand, as the merchant
increases $z$, the required mining power increases rapidly for low
$v$. For example, an attack with 1M~BTC goods at risk
(purple curve) can be successful with relatively low mining power,
$q = 0.25$, as long as the merchant keeps $z$ less than 25. But a
lower value $v$, such as 1K~BTC (yellow curve), would require the
attacker to possess mining power $q = 0.42$ for the same value~$z$.
\begin{figure}[t]  
\begin{minipage}{.7\columnwidth}
\centerline{\includegraphics[width=\columnwidth]{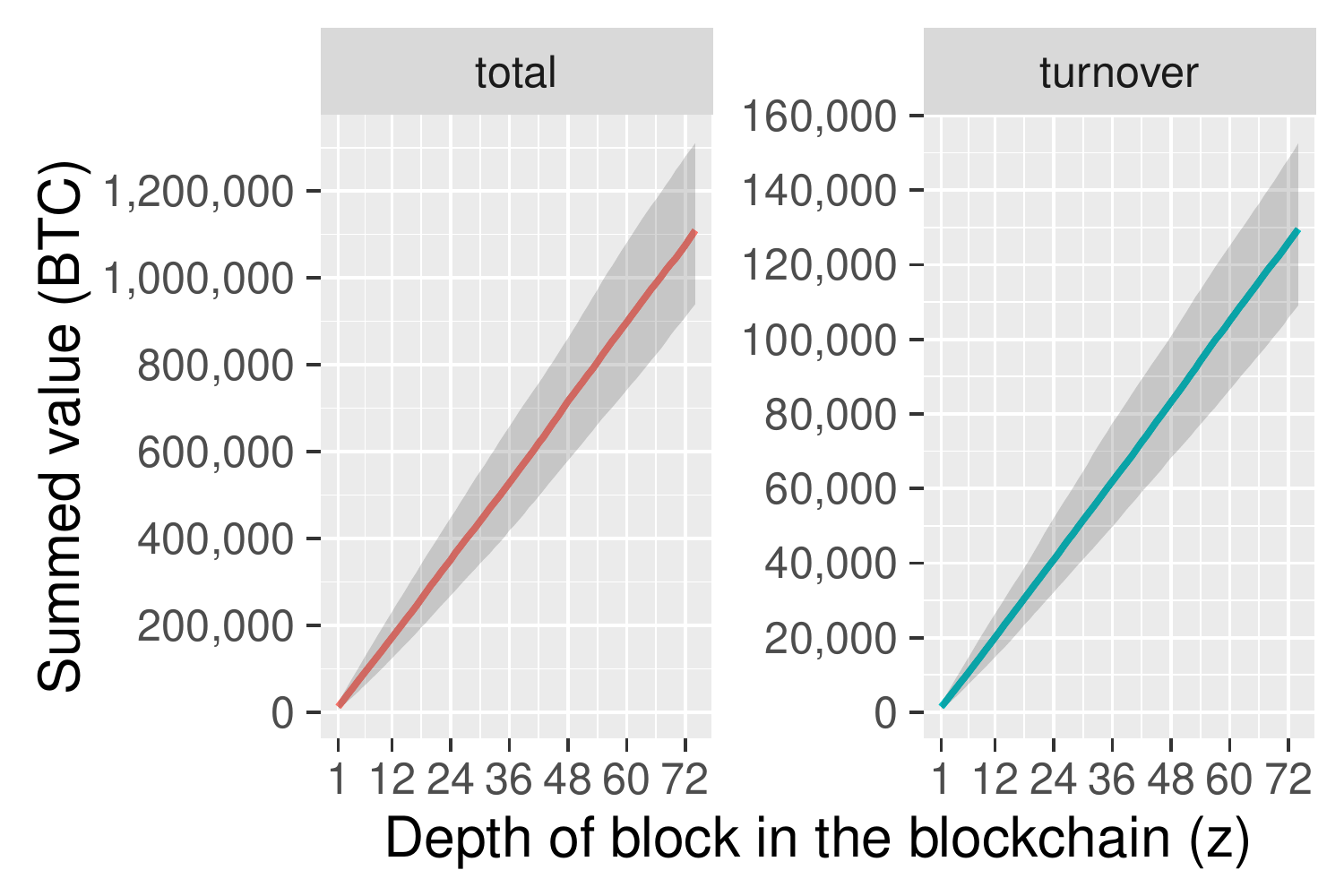}}
\end{minipage}\hfill
\begin{minipage}[c]{.29\columnwidth}
\caption{Total (red) and turnover (blue) block
  values (Bitcoin block data during July
  2016).  Each point on either plot corresponds to the
  median aggregate block value across a
  sliding window of $z$ consecutive blocks throughout the month. The shaded area represents the semi-interquartile range.}
\label{fig:min-q-blocks}
\end{minipage}
\end{figure}

\para{Determining value of goods at risk.} The attacker's potential profit has a strong
impact on his break-even point, and that profit is directly related to the goods at risk $v$. Thus, it is imperative that the merchant understand the scope of the attack. If she is confident that a potential attacker will target her alone, then the goods at risk can safely be assumed to be equal to the value of the goods she is personally trading for coin. On the other hand, if she would like to be conservative, then the merchant must assume that the attacker is capable of capturing the aggregate turnover in all confirmation blocks, which is the $z$-maximal goods at risk. 

Fig.~\ref{fig:min-q-blocks} shows the actual median aggregate total and turnover values for consecutively mined blocks during the
month of July 2016 (data source: blockchain.info).  Each point should be thought of as the typical
sum of all outputs for the transactions  or typical turnover value for the given number of consecutive blocks.  The plot shows that the merchant can significantly bound the
attacker's potential revenue by measuring turnover as opposed to using
total block output value.  Aggregate turnover values can be used in
conjunction with Fig.~\ref{fig:min-q-static-v} to determine the
merchant's security against an attacker who is capable of capturing the $z$-maximal goods. For example, if the merchant requires $z=6$ confirmations, and the observed aggregate $z$-block turnover is similar to the median aggregate turnover of 10K BTC, then she can be confident that she is protected against double-spends from an attacker with $q \leq 0.18$.

\para{Attacks by new miners.} Thus far we have assumed that the attacker is a \emph{defecting} miner, but it is also possible that a third party may bring new mining power to these attacks (purchased or, for example, stolen via a botnet), which we call \emph{expansion}. Recall that we assume the hash difficulty remains the same during the attack, so regardless of how the attacker garners mining power, he is always capable of mining the same expected number of blocks in a given period of time. An attacker who expands the mining power does not change our earlier analysis in Case 1 because the attacker is mining on a fraudulent branch that will never actually be publicly released. Therefore the salient factors, opportunity cost and mining rate, do not change. However the analysis does change for Case 2 because the attacker must compete with the other miners to grow his fraudulent branch longer than the main branch. For example, when an attacker with 1/3 of the existing mining power \emph{defects}, he can expect to mine approximately $\frac{\sfrac13}{\sfrac23}=\sfrac12$ of all mined blocks. In contrast, when \emph{expanding} the mining power by 1/3, the attacker only expects to mine blocks at approximately $\frac{\sfrac13}{1+\sfrac13}=\sfrac49$ the rate of the rest of the miners. Our existing analysis models the former scenario, but can easily be adjusted to model the latter by changing the honest miner block creation rate from $\beta_M = \sfrac{10}{(1-q)}$ to $\beta_M = 10$.

\para{Coinbase and fees.} Bitcoin's security is critically related to
the reward for mining. On July 9, 2016, the rewarded coinbase
halved from $B=25$ to $B=12.5$. All of our analysis assumes that $B=12.5$. In 2020, Bitcoin's security against
double-spends will decrease further since the coinbase reward will halve again.
A lower block reward absent higher fees or a significant increase in Bitcoin's fiat-exchange value will make it cheaper for attackers to procure a higher percentage of the mining power. Hence, if conditions remain the same except for a decrease in coinbase reward, then merchants will need to wait for more confirmations before releasing goods.

\para{Setting the deadline.} Our mitigation measures recommend that the merchant set a deadline $d$ when she suspects that there is risk of a concurrent eclipse and double-spend attack. Lower values of $d$ increase the break-even value for the attacker, which increases security.  The downside to securing goods with a shorter deadline is that honest customers may not succeed in meeting the deadline due to the inherent randomness of block discovery.  For example, according to actual block mining data we collected from 2016, only about 60\% of consecutive blocks of length $z$ actually arrived within a deadline of $d=10z$ minutes. Therefore, it is wise for the merchant and customer to agree on a contingency plan for cases where the deadline is missed. In many cases the customer will trust the merchant to issue a refund. In more adversarial settings, a third-party escrow service can be used to enforce a fair exchange of coin before the goods are released~\cite{Pagnia:2003}. If neither solution is acceptable, then the deadline should be relaxed so that the chances the entire mining community miss it are very small.

\para{Advice to \bc merchants.} Because it is impossible to know how many merchants will be targeted simultaneously, we recommend that merchants always choose $v$ equal to the $z$-maximal goods they observe for the $z$ blocks beginning with the one that confirms their transaction. Merchants are best off setting $z$ to address one of two cases. (1) Least conservatively, merchants could trust powerful miners to not carry out double-spend attacks, and assume that the common case is an attacker that has $q\leq0.1$ of the mining power. In this case, $z=1$ thwarts attackers when goods at risk are below 100~BTC, even under an eclipse attack. (2) Very conservatively, merchants can require $z=55$ confirmations (a little over 9 hours), which protects them against an attacker controlling $q \leq 0.35$ of the mining power even with goods at risk worth as much as 1M BTC (currently about \$700M).

\para{Applications to off-chain and side-chain protocols.} In its full generality, our analysis quantifies the security of an exchange of some off-blockchain quantity for Bitcoin. For most of this section, we have imagined a merchant trading physical goods or services. But our results apply equally to many systems that rely on or assume a stable blockchain, including sidechains~\cite{Back:2014}, micropayment channels such as the Lightning Network~\cite{Poon:2015} and TumbleBit~\cite{Heilman:2016},  and the XIM decentralized mix service~\cite{bissias:2014}.

All these alternate protocols require a certain transaction, $\cal{T}$, be confirmed in a Bitcoin block that locks or moves coin while in use by the other protocol. If the Bitcoin miners subsequently switch to a branch that doesn't include $\cal{T}$ (or worse, includes a transaction that prohibits $\cal{T}$'s validity in future blocks), then a {\em reorganization} of the alternate protocol's blockchain or transaction  is required, resulting in havoc. Unfortunately, due to the FLP impossibility result~\cite{Fischer:1985}, it is always possible for an attacker with sufficient resources to force a reorganization. And so in all cases, these protocols vaguely recommend the block containing the transaction reach a sufficient depth. For example, Back et al~\cite{Back:2014} recommends that ``a typical confirmation period would be on the order of a day or two.'' Sasson~\cite{sasson:2014} recommends that users with ``sensitive transactions only spend coins relative to blocks further back in the ledger''.

Using our analysis, the confirmation depth required for $\cal{T}$ can be more precisely calculated, and the risk of the reorganization can be quantified as follows. First, participants wait for $\cal{T}$ to be confirmed in a block. 
Then as confirmations accumulate, participants can use a resource like figure Fig.~\ref{fig:min-q-static-v} (which can be constructed for arbitrary parameter combinations using Eqs.~\ref{eq:c1-v} and \ref{eq:c2-v}) to determine the their security (in terms of attacker mining power $q$) given the current goods at risk. Eventually, they should settle on $z$ confirmations such that the security is considered to be sufficiently high for their purposes.

\section{Conclusion}\label{sec:conc}
We have presented a novel economic  model of \bc double spend attacks that incorporates the depth of the block containing the transaction of interest, the attacker's
mining power, goods at risk, and coinbase reward. Based on this model,
we have shown that the security of a transaction increases
roughly logarithmically with the number of confirmations that it receives, where an attacker
benefits from the increasing goods at risk but is also throttled by the
increasing proof of work required. Additionally, we have demonstrated
that, if merchants impose a conservative confirmation deadline, the eclipse attack does not increase an attacker's profit when his share of the mining power is less than 35\% or more than 10 confirmations are required.

\bibliographystyle{splncs03}
\bibliography{references}
\begin{figure*}[t]
\includegraphics[width=.32\columnwidth]{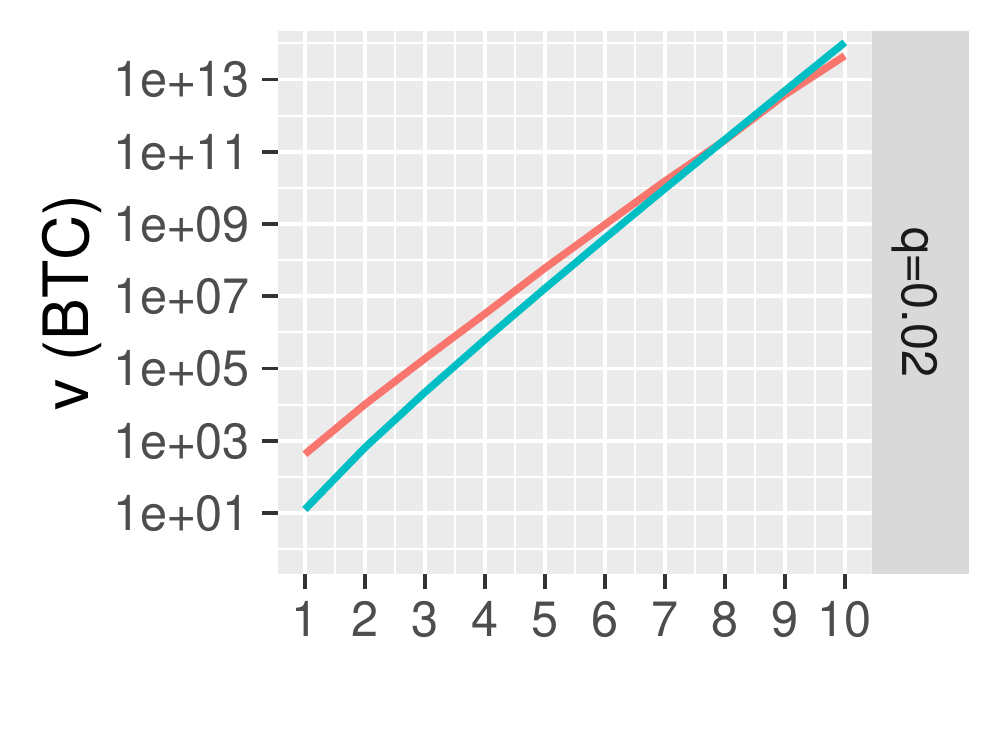}
\includegraphics[width=.32\columnwidth]{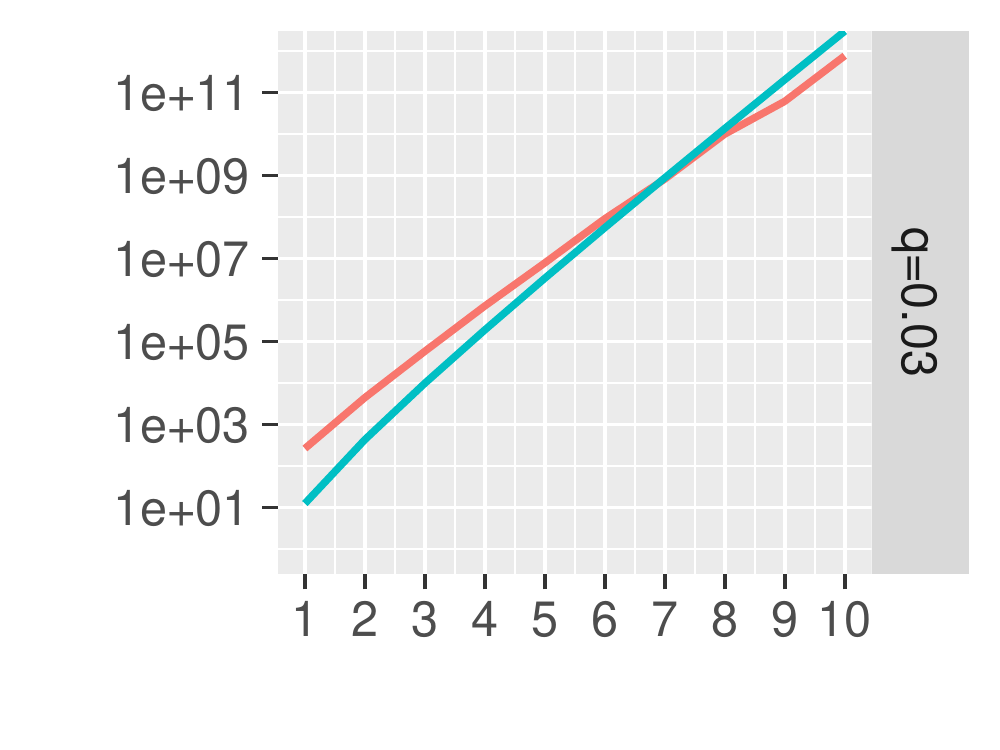}
\includegraphics[width=.32\columnwidth]{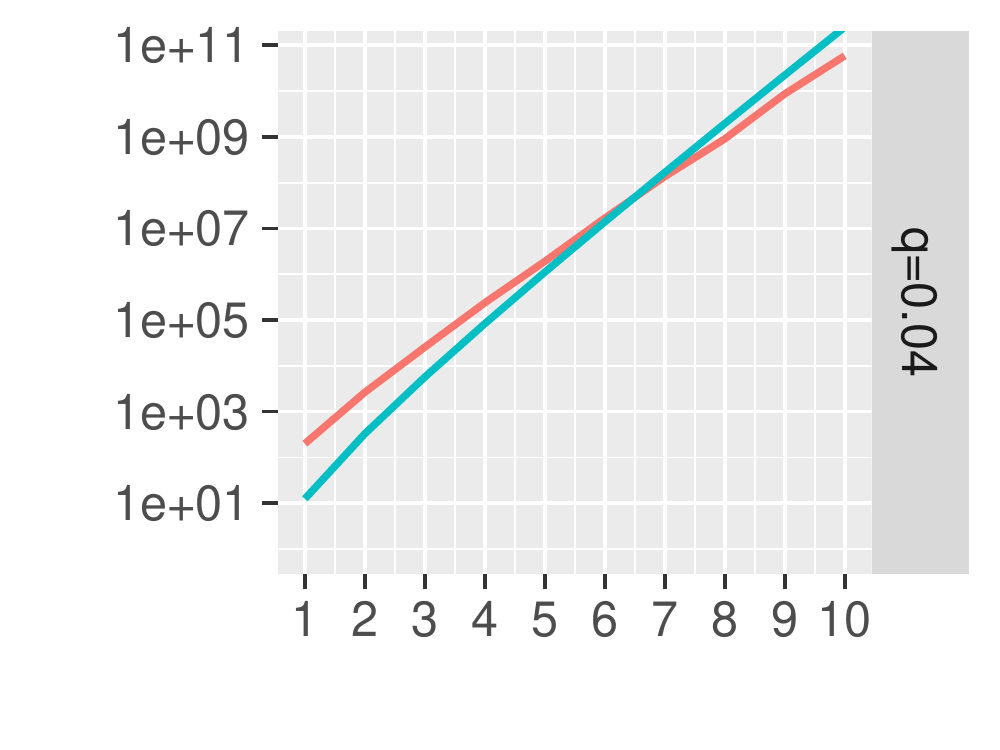}\\
\includegraphics[width=.32\columnwidth]{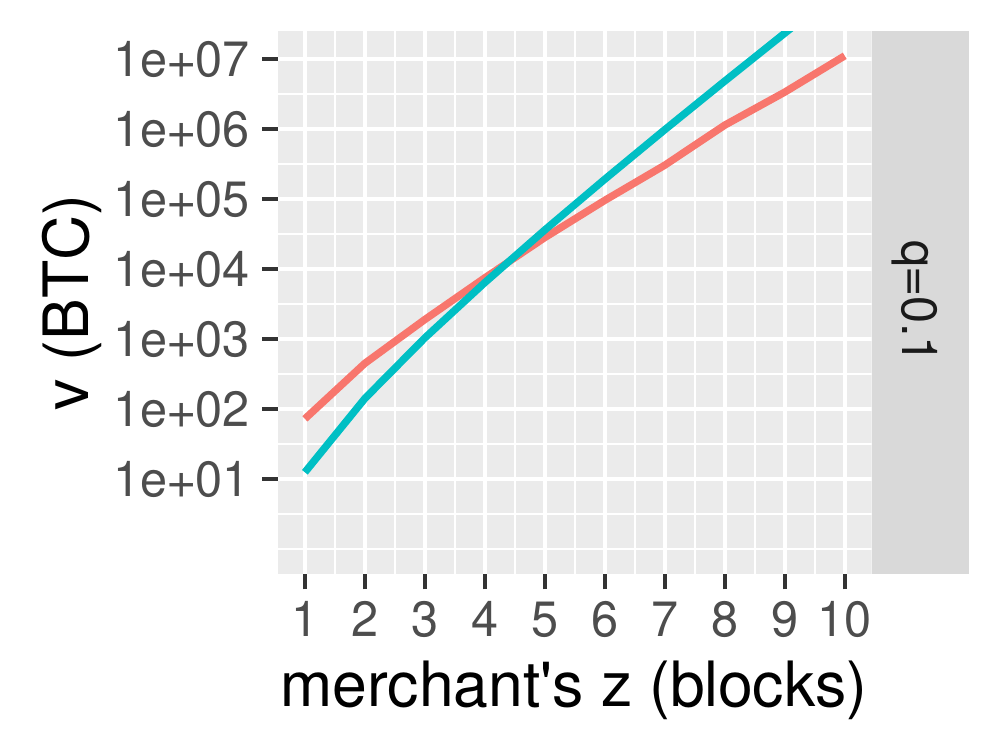}
\includegraphics[width=.32\columnwidth]{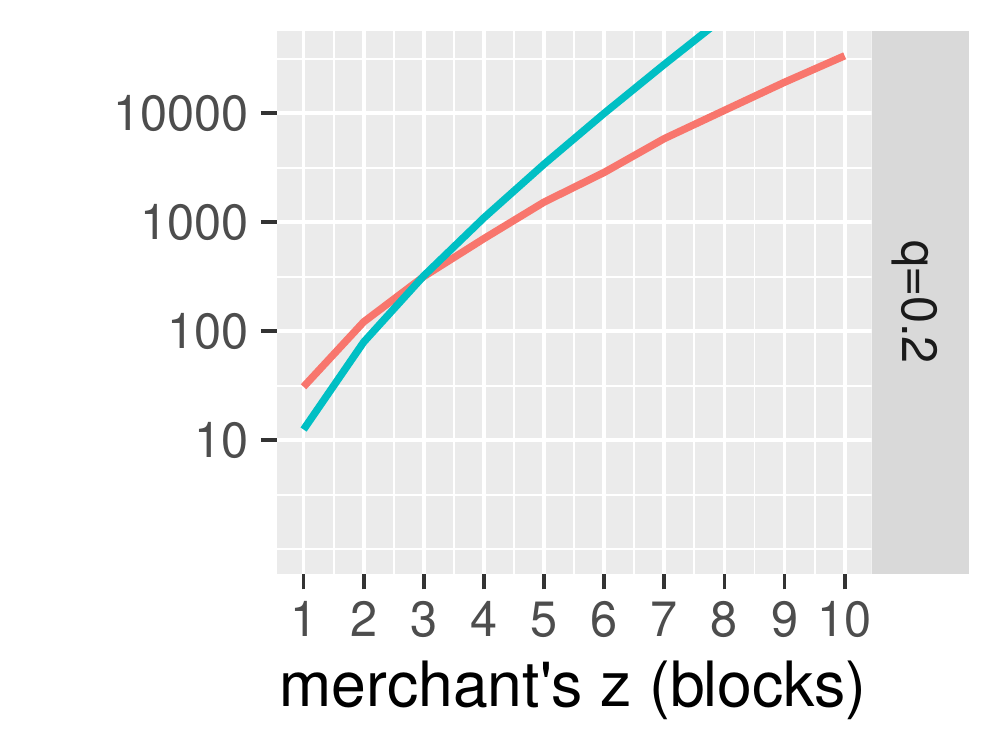}
\includegraphics[width=.32\columnwidth]{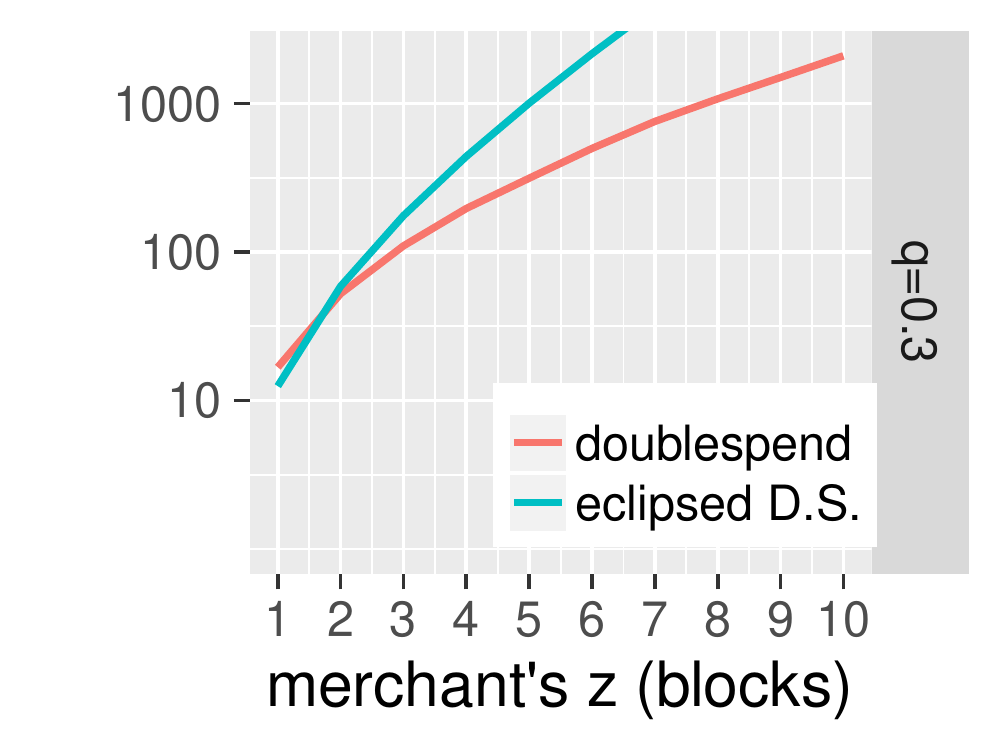}
\caption{An alternate visualization of Case 1 (Eq.~\ref{eq:c1-v}) and Case 2 (Eq.~\ref{eq:c2-v}) using a logscale on $y$ axis. }\label{fig:log}
\end{figure*}

\appendix
\section{Blockchain and Bitcoin Overview}\label{sec:backappendix}

The mining process places peers with significant computational power at an advantage, but overall, incentivizes miners to reach consensus.
\bc and follow-on currencies, such as Litecoin\cite{litecoin}, Zerocash~\cite{sasson:2014}, Ethereum\cite{ethereum}, and many others~\cite{bonneau:2015}, also use the blockchain algorithm to manage an electronic payment system.

Bitcoins exist as balance in a set of accounts called \emph{addresses}. \bc users exchange money through {\em transactions}, which transfer Bitcoin from one set of address to another. Transactions are broadcast over Bitcoin's p2p network where they are picked up by miners. Miners each independently agglomerate a set of transactions into a {\em block}, verify that the transactions are valid, and attempt to solve a predefined proof-of-work problem involving this block and all prior valid blocks. In Bitcoin, this process is dynamically calibrated to take approximately ten minutes per block.  Under ordinary, non-adversarial, conditions, the first miner to solve the proof-of-work problem broadcasts her solution to the network, adding it to the ever-growing {\em blockchain}; the miners then start over, trying to add a new block containing the set of transactions that were not previously added. 

A transaction appearing in a block is considered {\em confirmed}. When $z-1$ blocks have been added after the confirming block, the transaction is said to have received $z$ \emph{confirmations}. If two miners discover a new block simultaneously, the blockchain will bifurcate.  Miners will attempt to add to the branch with greatest cumulative proof-of-work. All miners will subsequently switch to mining on the first branch that grows longer. As incentive, all miners insert, as the first item in their block, a {\em coinbase} transaction, which is the
protocol-defined creation of new coins and a transfer of those coins into a address of
their choosing. In doing so, they have  {\em mined} those coins and made the chosen
address and its balance valid in future transactions. Miners also receive a small fraction of the face value of all transactions in the block that they successfully add to the blockchain; this transaction fee overhead serves to incentivize miners even after the last protocol-defined bitcoin is mined.
Miners are commonly organized into \emph{mining pools}, which allow many miners to pool together their resources. In these pools,  rewards are split equitably according to the amount of resources they contributed to creating a block.

\end{document}